\documentclass[twocolumn,showpacs,preprintnumbers,amsmath,amssymb]{revtex4}
\usepackage{epsfig}
\usepackage{graphicx}
\usepackage{epsfig}
\usepackage{color}
\usepackage{rotating}
\usepackage{amssymb}
\usepackage{amsmath}
\usepackage{graphicx}

\usepackage{amstext}
\usepackage{amsfonts}
\usepackage{amssymb,amscd}
\usepackage{amsopn}
\usepackage{graphicx}
\usepackage{bm,bbm}
\usepackage{epsfig,epic}

\def\beq{\begin{eqnarray}}
\def\eeq{\end{eqnarray}}
\def\la{\langle}
\def\ra{\rangle}

\newcommand{\be}{\begin{equation}}
\newcommand{\ee}{\end{equation}}
\newcommand{\bea}{\begin{eqnarray}}
\newcommand{\eea}{\end{eqnarray}}

\begin{document}

\title{Time reversals of irreversible quantum maps}

\author{Erik Aurell$^{1,2}$}
\email{eaurell@kth.se}
\author{Jakub Zakrzewski$^{3,4}$}
 \email{kuba@if.uj.edu.pl}
\author{Karol \.{Z}yczkowski$^{3,4,5}$}
\email{karol@tatry.if.uj.edu.pl}

\affiliation{
\mbox{$^1$ Dept. of Computational Biology and ACCESS Linnaeus Centre and Center for Quantum Materials,}
\mbox{KTH -- Royal Institute of Technology,   AlbaNova University Center, SE-106 91~Stockholm, Sweden}
\mbox{$^2$Depts. Information and Computer Science and Applied Physics and Aalto Science Institute (AScI), \\Aalto University, Espoo, Finland}
\mbox{$^3$ Instytut Fizyki imienia Mariana Smoluchowskiego, Uniwersytet Jagiello{\'n}ski, ulica {\L}ojasiewicza 11
30-348 Krak\'ow, Poland }
\mbox{$^4$ Mark Kac Complex Systems Research Center,
Uniwersytet Jagiello\'nski, Krak\'ow, Poland }
\mbox{$^5$ Center for Theoretical Physics, Polish Academy of Sciences, Warsaw, Poland}
}

\date{June 13, 2015}

\begin{abstract}
We propose an alternative notion of time reversal in open quantum systems as represented by linear quantum operations, and a 
related generalization of classical entropy production in the environment. 
This functional is the ratio of the probability to observe a transition between
two states under the forward and the time reversed dynamics, and leads, as in the classical case,
to fluctuation relations as tautological identities.
As in classical dynamics in contact with a heat bath, time reversal is not unique, and we discuss several possibilities.
For any bistochastic map its dual map preserves the trace and describes a legitimate 
dynamics reversed in time, in that case the entropy production in the environment
vanishes. For a generic stochastic map we construct a simple
quantum operation which can be interpreted as a time reversal.
For instance, the decaying channel, which sends
the excited state into the ground state with a certain probability,  
can be reversed into the channel transforming 
the ground state into the excited state with the same probability.
\end{abstract}


\maketitle

\textit{Introduction:}
The discovery of fluctuation relations~\cite{ES94,GC95,Jarzynski97}
has transformed classical out-of-equilibrium thermodynamics, giving rise to the new
field of stochastic thermodynamics
\cite{Searles08-review,Jarzynski11-review,seifert12-review,Sekimoto-book}. 
Important results obtained in the last two
decades include the use of Jarzynski equality to measure 
equilibrium free energy differences in large biomolecules from non-equilibrium measurements~\cite{GRB03,BLR05}, generalizations of fluctuation-dissipation theorems from the equilibrium to the non-equilibrium domain~\cite{CVdBK07,Gomez-Solano09}, and a sharpening of Landauer principle on the minimal heat generated in 
computing~\cite{B12,AGM12,Gawedzki13}. The central quantity in stochastic thermodynamics is the entropy production in the environment, a functional of the whole system 
history~\cite{Kurchan98,LS99} which can be defined in two ways.
The first method is by Clausius' relation 
\begin{equation}
\delta S_{env} = \beta \delta Q
\label{eq:deltaS-def-Clausius}
\end{equation}
where $\beta=\frac{1}{k_BT}$ is the inverse temperature and $\delta Q$ is the heat~\footnote{We here use the sign convention of
classical thermodynamics where heat is counted positive \textit{from} the system \textit{to} the bath.
In  stochastic thermodynamics literature the usual sign convention is
positive \textit{from} the bath \textit{to} the system~\cite{Sekimoto-book}.}
and the second way is as the Radon-Nikodym derivative of a forward and a reversed path probability
\begin{equation}
\delta S_{env}[\hbox{path}] = \log \frac{P^F[\hbox{path}]}{P^R[\hbox{path}]}.
\label{eq:deltaS-def}
\end{equation}
Here $P^F[\hbox{path}]$ is the probability of the forward path and $P^R[\hbox{path}]$ is the probability of the time-reversed path~\cite{ChG07},
and fluctuation relations follow from (\ref{eq:deltaS-def}) as mathematical ``tautologies''~\cite{Maes99,Gawedzki13}. 
Physically, fluctuation relations are, however, not tautologies, because the quantities in (\ref{eq:deltaS-def-Clausius}) and 
(\ref{eq:deltaS-def}) should be the same. For standard Markov models of the system-bath interactions (master equations, diffusion equations),
this is indeed the case, but for more general models the situation is less evident. 

The (possible) extension of fluctuation relations to the quantum domain has been the focus of intense
investigations reviewed in \cite{Esposito09-review, Campisi11-review}. Except for the 
generalizations of the Jarzynski equality and Crooks' fluctuation theorem to closed quantum 
systems~\cite{Kurchan-unpub} the results obtained to date lack the generality and simplicity of
fluctuation relations in classical systems, and typically hold for specific models such as
\textit{e.g.} when the quantum jump method~\cite{Legio13,HekkingPekola13,HorowitzParrondo13}
or the Lindblad formalism~\cite{ChetriteMallick} can be applied.

In this work we focus on the tautological aspect of quantum fluctuation relations, i.e. on the analogues
of (\ref{eq:deltaS-def}), which have not, we believe, been sufficiently emphasized in the literature.
To do this we have to define a general notion of time reversal of open quantum systems.
As in the case of a classical system interacting with a heat bath this notion of time reversal, which we call the $R$ operation, 
is not unique. In the sense introduced here an open quantum system can be time reversed in many
ways \cite{Cr08}, 
for each one one can define an entropy production functional analogous to (\ref{eq:deltaS-def}) and obtain fluctuation relations. 
We first give a very general (permissive) definition of the $R$ operation, and show that it always leads to fluctuation relations. We then turn to a possible definition of $R$ starting with the standard quantum mechanical time inversion of a combined system and the bath
and continuing to intrinsic representations in terms of Kraus operators. We discuss several special cases such as unital maps,
and give examples of time-reversals of 1-qubit channels.

\textit{Generalities and definition of $R$ operation:}
A state of a quantum system is described by its density matrix $\rho$ which 
is an $N$-dimensional positive Hermitian operator with unit trace.
Any physical operation on a quantum system can be described by a completely positive 
linear map $\Phi$,
which preserves the trace and 
sends one density matrix $\rho$ into another state $\rho'=\Phi(\rho)$. 
We define a general 
time reversal $R$ as an involution on the set $\Omega_N$ of quantum operations 
\textit{i.e.} a bijective transformation $R:\Phi\to\Phi^R$ which satisfies $(\Phi^R)^R=\Phi$.

\textit{$R$ operation and fluctuation relations:}
To arrive at fluctuation relations 
we consider the paradigmatic example of
two measurements, one before the beginning of the process described by $\Phi$, and another after~\cite{Kurchan-unpub}. 
We will denote by $\hat A$ and $\hat O$  two measured operators
(with eigenstates $|a\ra$ and $|o\ra$ and eigenvalues $E_a$ and $E_o$).
At the beginning of the process the system will hence be in the
pure state $|a\ra\la a|$, and the probability to observe $o$ at the end will be 
$\la o|\Phi(|a\ra\la a|)|o\ra$.
For the reversed chain of events, we first measure $\hat O$ to obtain
$o$, then act on the system with the reversed quantum map $\Phi^R$, and measure 
$\hat A$ to obtain $a$.
This happens with probability $\la a|\Phi^R(|o\ra\la o|)|a\ra$. A straightforward generalization of
(\ref{eq:deltaS-def}) to the quantum domain is thus
\begin{equation}
\delta S_{env}[a,o] = \log \frac{\la o|\Phi(|a\ra\la a|)|o\ra}{\la a|\Phi^R(|o\ra\la o|)|a\ra}
\label{eq:qdeltaS-def}
\end{equation} 
To derive fluctuation relations from (\ref{eq:qdeltaS-def}) one proceeds by analogy with the classical case using (\ref{eq:deltaS-def}).
For Jarzynski equality one takes $\hat A$ as an initial Hamiltonian $\hat H_i$ and 
$\hat O$ as a final Hamiltonian $\hat H_f$
with thermodynamic equilibrium states $\rho^{\beta}_i$ and $\rho^{\beta}_f$, respectively, 
and defines the work done on the system during the process as 
$\delta W[a,o] = \left(E_o-E_a\right)+\beta^{-1}\delta S_{env}[a,o]$. 
Then $\sum_{a,o} e^{-\beta E_o} \la a|\Phi^R(|o\ra\la o|)|a\ra = Z_f$ 
and this identity can simply be rewritten as
$$\sum_{a,o} \langle a| \rho^{\beta}_i|a\rangle\langle o|\Phi(|a\rangle\langle a|)|o\rangle e^{-\delta S_{env}{a,o}} e^{-\beta (E_o-E_a)}=e^{-\beta \Delta F}.$$
Here $\Delta F$ denotes the difference between the free energy of both thermal equilibrium states, and the left hand side
can be interpreted as $\langle e^{-\beta\delta W}\rangle$.
For Crooks' relation one similarly defines
$P^F(x) = \la\delta\left(\delta W-x\right)\ra_{eq}$
and  $P^R(x) = \la\delta\left(\delta W^R-x\right)\ra^R_{eq}$ where in the latter 
the time-reversed work is $\delta W^R[o,a]=-\delta W[a,o]$ and 
the average is performed over the final equilibrium state $\rho^{\beta}_f$ and the 
reversed quantum map $\Phi^R$.
This implies $P^R(-x)=e^{-\beta \left(x-\Delta F\right)}P^F(x)$, which is the relation of Crooks \cite{Cr98}.

\textit{Standard quantum mechanical time inversion and time reversal of environmental representations}
Consider first  a closed system.
If $\rho'=\Phi\rho=U\rho U^{\dagger}$ where $U={\cal T}e^{-\frac{i}{\hbar}(t_f-t_i)H}$ is unitary 
development with Hamiltonian $H$ then the final state is time inverted by an 
antiunitary operator $\Theta$ \textit{i.e.} $\tilde{\rho'}=\Theta\rho'\Theta^{-1}$.
Acting on this state with $\tilde{U}=\Theta U^{\dagger}\Theta^{-1}$,
we obtain a state at the initial time $\tilde{\rho}=\tilde{U}\tilde{\rho'}\tilde{U}^{\dagger}$
and time inversion is complete if $\rho=\Theta^{-1}\tilde{\rho}\Theta$, which is the case.
The time reversal of a closed system, based on standard quantum mechanical time inversion, is then
the operation $\Phi\to\tilde{\Phi}$ where  $\tilde{\Phi}\rho=\tilde{U}\rho\tilde{U}^{\dagger}$.
Obviously this is an involution: performing it  twice we get back the map $\Phi\rho$.

Quantum maps have (generally non-unique) \textit{environmental representations}
\begin{equation}
\rho'=\Phi\rho=\hbox{Tr}_B\left[U\left(\rho\otimes \sigma\right)U^{\dagger}\right]
\label{environment}
\end{equation}
where the principal state $\rho$ acts in the space ${\cal H}_A$
while the ancillary state $\sigma$ acts in the space ${\cal H}_B$
describing the environment. Both subsystems are coupled by a unitary operation $U$,
and the image $\rho'$ is obtained by tracing out the environment.
Under standard quantum mechanical time inversion $U$ transforms to $\tilde{U}$ and
$\sigma$ to $\tilde{\sigma}$ which allows us to define an \textit{environmental} time 
reversal of the open quantum system as
\begin{equation}
\Phi^{R_E}\rho=\hbox{Tr}_B\left[\tilde{U}\left(\rho\otimes \tilde{\sigma}\right)\tilde{U}^{\dagger}\right]
\label{environment-R}
\end{equation}
The disadvantages of such a definition are obvious: it is contingent on the chosen environmental representation,
and to arrive at an intrinsic notion we have to pursue another approach. In addition, also when an open quantum operation
has a very natural environmental representation, (\ref{environment-R}) is not the only natural notion
of generalized time reversal. We will return to this point below.

\textit{Kraus form of quantum maps and the dual map:}
Instead of (\ref{environment-R}) we will now start from the {\sl Kraus form}  \cite{Kr71} of quantum operator $\Phi$:
\begin{equation}
\rho'=\Phi(\rho)=\sum_{i=1}^k  A_i \rho A_i^{\dagger} .
\label{Kraus}
\end{equation}
As a quantum operation $\Phi$ preserves the trace, Tr$\rho'={\rm Tr}\rho$,
the Kraus operators $A_i$ satisfy the 
identity resolution $\sum_{i=1}^k A_i^{\dagger} A_i = {\mathbbm 1}$.
The trace preserving conditions induce $N^2$ constraints
 so the set $\Omega_N$ of quantum operations acting on $N$
dimensional states has $N^4-N^2$ dimensions.
Although the number $k$ in (\ref{Kraus}) is arbitrary, for any map 
there exist the {\sl canonical Kraus form}, 
for which all Kraus operators are orthogonal,
\begin{equation}
{\rm Tr} A_iA_j^{\dagger}=d_i \delta_{i,j} ,
\label{ortho}
\end{equation}
so that the Kraus rank $k\le N^2$.
In a generic case the numbers $d_i$ are different
and this representation is unique
up to the choice of the overall phases of the Kraus operators --
see \textit{e.g.}~\cite{BengtssonZyczkowski}.

In the case of a unitary evolution, 
$\rho'=\Psi_U(\rho)=U\rho U^{\dagger}$,
one has $k=1$  and the only Kraus operator is unitary, $A_1=U$.
Unitary maps belong to broader class of {\sl unital maps},
which preserve the identity, $\Psi({\mathbbm 1})={\mathbbm 1}$,
and satisfy the dual condition
$\sum_{i=1}^k A_i A_i^{\dagger}= {\mathbbm 1}$.
A map which is trace preserving and unital is called 
{\sl bistochastic},
as it forms a quantum analogue of a bistochastic matrix,
which acts in the set of $N$--point probability vectors.

For any quantum map $\Phi$ in the form (\ref{Kraus})
one defines its {\sl dual map} $\Phi^{\circ}$
such that $\Phi^{\circ}(\rho)=\sum_{i=1}^k A_i^{\dagger} \rho A_i$.
Writing for short $\Phi=\{A_1, \dots, A_k\}$
we have $\Phi^{\circ}=\{A_1^{\dagger}, \dots , A_k^{\dagger}\}$.
Note that if a map $\Phi$ preserves the trace, the dual map 
$\Phi^{\circ}$ is unital. If $\Phi$ is bistochastic, so is
$\Phi^{\circ}$. It is also convenient to define the class of
{\sl selfdual maps} which satisfy $\Phi^{\circ}=\Phi$.
A quantum operation for which all Kraus operators 
are hermitian is bistochastic and selfdual.
So is a map for which all non-hermitian operators occur in pairs, 
e.g. $\Phi=\{A_1, \dots, A_m, A_1^{\dagger}, \dots, A_m^{\dagger} \}$.

Restricting our attention to bistochastic maps as a first generalization of the standard time inversion,
we can  define a time reversal $\Phi^{R_B}:=\Phi^{\circ}$.
The choice $R_{B}$ implies that
$\la o|\Phi(|a\ra\la a|)|o\ra= \sum_{i}|\langle o |A_i | a\ra|^2= 
\la a|\Phi^R(|o\ra\la o|)|a\ra$. Therefore the fraction in Eq.~(\ref{eq:qdeltaS-def}) is equal to unity
and hence for any bistochastic map $\Phi$ and any initial and final pure states
the entropy production vanishes, $\delta S_{env}[a,o]=0$.
Although bistochastic maps are not time invertible in the standard sense, they 
are therefore time reversible in the sense that there 
exists a (natural) time reversal of such maps for which the entropy production in the environment vanishes.
As a consequence fluctuation relations hold for all such maps with the work taken equal to the internal energy change
which reproduces the original quantum fluctuation relation of Kurchan for unitary maps~\cite{Kurchan-unpub} 
as well as the result of Rastegin for bistochastic maps~\cite{Rast13}. 
In this connection it was recently observed  \cite{Alb,RZ14} that 
the Jarzynski relation, with work taken equal to internal energy change,
in general does not hold for nonunital maps.
 Thus if the map is nonunital then 
its dual is not a well-defined quantum map, time reversal cannot be defined
as a dual, and the ratio in Eq. (\ref{eq:qdeltaS-def}) may be different from unity.

A more general definition of the time reversal operation for a non-unital a quantum operation
$\Phi=\{A_1, \dots, A_k\}$  was proposed by Crooks \cite{Cr08}.
It is based on the invariant state of the map, 
$\rho_*=\Phi(\rho_*)$. 
The reversed map is given by the following sequence of the Kraus operators
$\Phi^{R_C}=\{A^C_1, \dots, A^C_k\}$, 
where  ${A^C_i}= \rho_*^{1/2} A_i^{\dagger} \rho_*^{-1/2}$. 
Note that in the case of unital maps one has 
$\rho_*^{1/2} ={\mathbbm 1}/N$ so that $A^C_i= A_i^{\dagger}$
and one arrives at the dual map, $\Phi^{R_C}= \Phi^{\circ}$.
However, this definition does not work, if the invariant state 
 belongs to the boundary of the set of quantum states so that $\rho_*$ is not invertible.

The isomorphism of Choi and Jamio{\l}kowski \cite{Cho75a,Ja72,BengtssonZyczkowski}
states that an operation $\Phi$ can be uniquely described by 
a {\sl dynamical matrix}, (Choi matrix)
$$D_{\Phi}=N (\Phi \otimes {\mathbbm 1}) |\psi^+\rangle \langle \psi^+|,$$
where $|\psi^+\rangle=\frac{1}{\sqrt{N}} \sum_{i=1}^N |i\rangle \otimes |i\rangle$
denotes the maximally entangled state on an extended Hilbert space
${\cal H}_A \otimes {\cal H}_B$.
The matrix $D$ of order $N^2$ is hermitian and positive by construction,
and its eigenvalues determine the relative weights $d_i$,
while the eigenvectors of length $N^2$, reshaped into square matrices of 
order $N$ and rescaled by $\sqrt{d_i}$ yield the Kraus 
operators $A_i$ in the canonical form (\ref{ortho}).

Two maps $\Phi_1$ and $\Phi_2$ are called {\sl unitarily equivalent},
written $\Phi_1 \sim \Phi_2$,
if there exist two unitary matrices $V_1$ and $V_2$ such that
 $\Phi_2(\rho)= V_2\bigl(\Phi_1(V_1 \rho V_1^{\dagger})\bigr)V_2^{\dagger}$, 
so the map $\Phi_2$ can be written as a concatenation of $\Phi_1$ 
with two unitary operations,
$\Phi_2= \Psi_{V_2} \circ \Phi_{1} \circ \Psi_{V_1}$.
Observe that for any two unitarily equivalent maps
the corresponding dynamical matrices, $D_1$ and $D_2$,
share the same spectrum and are unitarily similar.

\textit{Essential map and its time reversal:}
We now look for a possible choice  of the involution $R$
for a general, non--unital quantum map, for which $\Phi^{\circ} \notin {\Omega_N}$
Consider a generic quantum operation $\Phi$, for which the
spectrum $\{d_i\}_{i=1}^k$ 
of the corresponding dynamical matrix $D_{\Phi}$ is non-degenerate.
The trace of the dynamical matrix is fixed, Tr$D=\sum_{i=1}^k d_i$, 
so let us order the Kraus operators forming the canonical form
(\ref{ortho}) according to their norms, 
$d_1=||A_1||^2 \ge d_2 =||A_2||^2 \ge , \dots, \ge d_k=||A_k||^2$.
The leading Kraus operator, with the largest norm,
is represented by a possibly non-hermitian matrix $A_1$ of order $N$,
which can be brought to the diagonal form by the singular value decomposition,
\begin{equation}
  A_1 = V_1 E V_2^{\dagger}, {\rm \ \ where \ \ } V_1V_1^{\dagger}=V_2V_2^{\dagger}={\mathbbm 1} .
\label{svd1}
\end{equation}
Here $E$ is a diagonal matrix with all non-negative entries.
In the generic case the spectrum of $E$ is non-degenerate,
and this decomposition is unique up to the phases of the right and left 
eigenvectors of $A_1$ which form unitary matrices $V_1$ and $V_2$. 
In the degenerate case this decomposition is
not unique. For instance, if $A_1=U$ is unitary, than $E={\mathbbm 1}$,
and one can choose e.g. $V_1=U$ and $V_2={\mathbbm 1}$.

For any map $\Phi$ we select in this way two unitary matrices $V_1$ and $V_2$,
which allow us to  define rotated Kraus operators $B_i=V_1^{\dagger} A_i V_2$ 
and the {\sl essential map} 
\begin{equation}
{\hat \Phi}(\rho)=\sum_{i=1}^k B_i \rho B_i^{\dagger} =  
    V_1^{\dagger} A_i  V_2 \rho V_2^{\dagger} A_i^{\dagger} V_1 .
\label{Kraus2}
\end{equation}
For any operation determined by a set of ordered Kraus operators,
$\Phi=\{A_1, A_2,\dots, A_{k} \}$ the corresponding
essential map reads thus 
$\hat{ \Phi}=\{E, V_1^{\dagger} A_2 V_2, \dots, V_1^{\dagger} A_k V_2 \}$,
where $A_1=V_1E V_2^{\dagger}$.
Observe that the map $\Phi$ and its essential map are unitarily equivalent,
\begin{equation}
 \Phi \sim \hat {\Phi} \ = \  \Psi_{V_1} \circ \Phi \circ \Psi_{V_2^{\dagger}}  .
\label{essen2}
\end{equation}
It is easy to see that for any unitary evolution $\Psi_U$ the corresponding essential
map reduces to identity map, $\hat{\Psi}_U={\mathbbm 1}$. Thus the essential map
generically provides a unique description of non-unitary part of a discrete quantum evolution.

Consider, for instance a one--qubit {\sl Pauli channel}, 
\begin{equation}
\Phi_p(\rho) = \sum_{j=1}^4 p_j A_j \rho A_j,
\label{Paul2}
\end{equation}
where  $p_j$ is an ordered,  $p_1\ge p_2\ge p_3 \ge p_4$,
normalized probability vector, $\sum_{j=1}^4 p_i=1$,
while hermitian Kraus operators $A_j$ form
an arbitrary sequence of three Pauli matrices $\sigma_i$, $i=1,2,3$ 
and identity,  $\sigma_0={\mathbbm 1}_2$.
Then the corresponding essential map reads 
\begin{equation}
{\hat \Phi_p}(\rho) = p_0 \rho + \sum_{i=1}^3 q_i \sigma_i \rho \sigma_i,
\label{Paul3}
\end{equation}
where the three coefficients $q_i$ are up to a permutation  equal up to
the three smallest coefficients $p_2, p_3, p_4$ of the map $\Phi$.

In other words any Pauli channel can be represented by a point in the regular 
simplex $\Delta_3  \subset {\mathbbm R}^3$
 spanned by the identity and Pauli matrices (see Fig. 1a).
The corresponding essential map belongs then to the asymmetric fourth part of the
simplex with a corner representing the identity map ${\mathbbm 1}$.

\begin{figure}[htbp]
        \centerline{ \hbox{
              \epsfig{figure=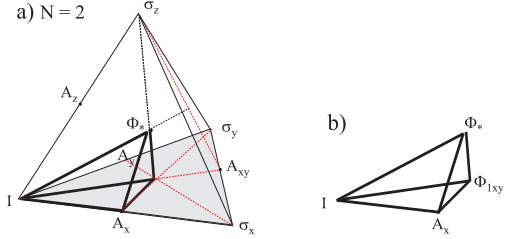,width=7.0cm}
                   }}
\caption{a) The set of one-qubit Pauli channels forms a probability simplex  $\Delta_3$
spanned by identity and Pauli matrices. b) An asymmetric part of the simplex
forms a fragment of the set of essential maps ${\hat \Phi_p}$ which 
contains the identity map  ${\mathbbm 1}$ and the maximally depolarizing 
channel ${\Phi_*}$. Channel $A_{xy}$ forms the symmetric combination
of  $\sigma_x$ and $\sigma_y$.
}
        \label{fig1}
\end{figure}

{\it The essential map and an intrinsic $R$:}
Having defined an essential map $\hat \Phi$ , which describes the
non-unitary part of the evolution, 
$\Phi= \Psi_{V_1^{\dagger}} \circ {\hat \Phi} \circ \Psi_{V_2}$,
 we are in position to give an intrinsic definition of a {\sl time reversed map}
\begin{equation}
\Phi^R := \Psi_{V_2} \circ {\hat \Phi} \circ \Psi_{V_1^{\dagger}}.
\label{rev1}
\end{equation}
Looking at the reversed map in the Kraus representation
$\Phi^R=
\{A_1^{\dagger},  Y A_2  Y, \dots  Y A_k Y\}$
with the unitary matrix $Y=V_2 V_1^{\dagger}$,
we see that the leading Kraus operator $A_1$ is indeed inverted into $A_1^{\dagger}$,
while other operators are suitably rotated to keep the map $\Phi^R$ trace preserving.

Making use of the form (\ref{essen2}) we see that
the composition of a map with its reverse reads
\begin{equation}
\Phi^R \circ \Phi = \Psi_{V_2} \circ {\hat \Phi} \circ {\hat \Phi} \circ  \Psi_{V_2^{\dagger}},
\label{rev2}
\end{equation}
and is unitarily similar to ${\hat \Phi} \circ {\hat \Phi}={\hat \Phi}^2$.

It is easy to check that for any map the following properties hold true
$\widehat{(\Phi^R)}=\hat{\Phi}=(\hat{\Phi})^R$
and  $(\Phi^R)^R=\Phi$, so that the reverse operation $^R$ is an involution,
as requested.  If the map is unitary, $\Phi=\Psi_U$,
 such a composition reduces to the identity map
\begin{equation}
\Psi_U^R \circ \Psi_U = \Psi_{U^{\dagger}} \circ \Psi_{U} = {\mathbbm 1},
\label{rev3}
\end{equation}
as any unitary evolution can be reversed.

Note that for a selfdual map one has $\Phi^R=\Phi^{\circ}=\Phi$
so that $\Phi^R \Phi = \Phi^2$. For instance, consider the  selfdual
maximally depolarizing channel, which sends any state into maximally mixed state,
$\Phi_*(\sigma)={\mathbbm 1}/N$. For any two pure states one has
$|\langle i| \Phi_*(|j\rangle \langle j|) |i\rangle| =1/N$
so that the fraction in Eq.~(\ref{eq:deltaS-def}) is equal to unity
for any choice of initial and final states.

For any stochastic map described by two Kraus operators, $\Phi=\{ A_1, A_2\}$, 
the reversed map reads
$\Phi^R=\{A_1^{\dagger}, \sqrt{{\mathbbm 1} - A_1 A_1^{\dagger}} \}$.
In this case, the operators $A_i$ need not to be ordered according to their norms,
so one can choose for $A_1$ a non-hermitian operator.
For example, in the case of the one-qubit decaying channel,
$\Phi_{\rm dec}=\Bigl\{ 
{\small \left[\begin{array}{cc}
                  0  & \sqrt{p}  \\
                  0  & 0   \\
\end{array}\right] }, 
{\small  \left[\begin{array}{cc}
                  1  &  0  \\
                  0  & \sqrt{1-p}   \\ \end{array}\right]} 
 \Bigr\}$, 
where $p \in [0,1]$ is a free parameter,
the dual map $\Phi_{\rm dec}^{\circ}=\{A_1^{\dagger}, A_2^{\dagger} \}$
is not stochastic. However the inverted map
$\Phi_{\rm dec}^R=\Bigl\{ 
{\small \left[\begin{array}{cc}
                  0  &    0  \\
                  \sqrt{p}  & 0   \\
\end{array}\right] }, 
{\small  \left[\begin{array}{cc}
                  \sqrt{1-p}   &  0  \\
                  0  & 1    \\
\end{array}\right]} \Bigr\}$, 
is stochastic. While the decaying channel $\Phi_{\rm dec}$
describes the process of a spontaneous decay 'downwards' $|1\rangle \to |0\rangle$
with probability $p$,  the reversed process $\Phi_{\rm dec}^R$
describes the transition 'upwards'
$|0\rangle \to |1\rangle$. Observe that for such a definition
of the reversed map $\Phi_{\rm dec}^R$
the Crooks relation holds as a tautology.
Note also that the invariants states of the map 
$\Phi$ and its reverse $\Phi^R$ are different.
This is not the case for 
the time reversal operation $\Phi^{R_C}$ from \cite{Cr08}.
Furthermore, in the case $p>0$
the invariant state of the map $\Phi_{\rm dec}$
is pure and thus not invertible,
so the operation $\Phi_{\rm dec}^{R_C}$ is not well defined.

\textit{Reversing Quantum Brownian motion}
We now revert back to the environmental representation (\ref{environment}) and assume that the system actually
is connected to second physical system which acts as a heat bath. 
Such a map can be given by (\ref{environment}) where
the ancilla state $\sigma$ is a thermal equilibrium state of a set of harmonic oscillators at inverse temperature $\beta$, and $U$
is  a unitary time development of the system plus the bath determined by a total Hamiltonian $H=H_S+H_I+H_B$ where $H_S$ is the system Hamiltonian,
$H_B$ is the bath Hamiltonian and $H_I$ is a linear interaction of the system and the bath.
Standard time reversal of such a map is then given by
(\ref{environment-R}), a procedure that can here be described as ``attach the time-inversed bath, and run time backwards''.

The quantum Brownian motion model is based on the observation that if
bath frequencies form a continuum with Ohmic spectrum
and a spectral cut-off, and the temperature goes to infinity, then
(\ref{environment-R}) 
represents classical Kramers-Langevin dynamics 
$\dot{p}=f(x,t)-\eta\frac{p}{M} +\sqrt{2\eta/\beta}\dot{\xi}$ and $\dot{x}=\frac{p}{M}$~\cite{CaldeiraLeggett83,BreuerPetruccione}. 
Applying (\ref{environment-R}) to the quantum Brownian motion model all operators are time reversed according to their parity, 
which in the classical limit means $t\to t^*=t_f-t$, $x_t\to x^*_t+= x_{t^*}$ and $p_t\to p^*_t+= -p_{t^*}$, and 
the Kramers-Langevin equation transforms into $\frac{dp^*}{dt^*}=f(x^*,t_f-t^*)+\eta\frac{p^*}{M} +
\sqrt{2\eta/\beta}\dot{\xi}$ and $\frac{dx^*}{dt^*}=\frac{p^*}{M}.$
Time reversals of classical stochastic differential equations have 
been extensively discussed in the literature,
and it is well understood that they are not unique~\cite{ChG07}.
The example just derived by taking the classical limit of quantum Brownian motion is the
``natural time reversal'' of Kramers-Langevin dynamics, but also other possibilities make sense 
(note the ``anti-friction''!).

A second example of time reversal of Kramers-Langevin dynamics, in \cite{ChG07} called ``canonical time reversal'', is based on the same variable transformation but assumes that the conservative force and
the friction force transform differently under time inversion, resulting in
a Kramers-Langevin dynamics also for the time-reversed motion \textit{i.e.} 
$\frac{d}{dt^*}p^*=f(x^*,t_f-t^*)-\eta\frac{p^*}{M} +\sqrt{2\eta/\beta}\dot{\xi}$ and $\frac{d}{dt^*}x^*=\frac{p^*}{M}$.
To lift this definition to ``$R$'' we obviously have to treat the system and the bath differently.
For the bath time must run forwards, so as to result in dissipation, while the conservative effects embodied in the force $f$
are to be applied in the opposite order. This can be achieved by considering the system Hamiltonian of the form
$H_S=H_{kin}+V(x_S,t)$ and the two unitary operators
\begin{eqnarray}
U &=& {\cal T} e^{-\frac{i}{\hbar}\left(\int_{0}^{t_f}H_{kin}+V(x_S,t)+H_I+H_B\right)}\\
U^{R_C}&=&{\cal T} e^{-\frac{i}{\hbar}\left(\int_{0}^{t_f}H_{kin}+V(x_S,t_f-t)+H_I+H_B\right)}
\end{eqnarray}
where ${\cal T}$ stands for time ordering.
Time reversal by changing $U$ to $U^R$ amounts to the procedure of 
``attach the bath, let time run forwards, but time-reverse the 
external drive''. In \cite{ChG07} several other examples are given of time reversals of stochastic dynamics
which can also be ``lifted to $R$''.

\textit{Discussion:} In this work we have introduced a general notion of time reversals
of quantum maps which generalizes standard time inversion in quantum mechanics.
As in classical dynamics in contact with a heat bath, this definition of time reversal is not unique.
One possibility -- but not the only possibility -- is to choose an environmental representation of
the quantum map, and then apply standard quantum time inversion on the combined system and ancilla.
Another possibility is to start from an intrinsic definition of the quantum map in terms of Kraus
operators, and then define time reversal on that level. In any case, from any such definition one 
can define an entropy production in the environment functional analogously to the classical
setting, and for each such definition quantum fluctuation relations are satisfied identically.

\section*{Acknowledgements}
This research is supported by the Swedish Science Council through grant 621-2012-2982
and by the Academy of Finland through its Center of Excellence COIN (EA), National Science Centre (Poland)
through grants  DEC-2012/04/A/ST2/00088 (JZ) and DEC-2011/02/A/ST1/00119 (K\.Z) as well as EU FET project QUIC 641122 
and the project Focus KNOW at the Jagiellonian University.


{\bf Note added}. After the first version of this paper was posted in the arXiv
a related work \cite{MHP15} appeared.

\end{document}